\documentclass[a4paper,12pt]{article}
\usepackage{graphicx}
 \usepackage{mathptmx}
 \usepackage{amssymb}
 \usepackage{latexsym}
\def\dst{\displaystyle}
\begin{document}
\title{\bf Dispersion relations for meson-proton and proton-proton forward elastic scattering}
\author{A. Alkin \thanks{Bogolyubov Institute for Theoretical Physics,
             Metrolologichna 14b, Kiev, Ukraine, 03680} \thanks{email: alkin@bitp.kiev.ua} \and  J.R. Cudell \thanks{IFPA,
           AGO dept., University  of Li\'{e}ge, 4000 Li\'{e}ge, Belgium, email: jr.cudell@ulg.ac.be} \and E. Martynov$^{*}$ \thanks{email: martynov@bitp.kiev.ua}}

\maketitle

\begin{abstract}
An analysis of the data on forward $pp, \bar pp, \pi^{\pm}p$ and $K^{\pm}p$
scattering is performed making use of the single- and double-subtraction
integral and comparing with derivative dispersion relations for amplitudes.
Various pomeron and odderon models for the total cross sections are considered
and compared.
The real part of the amplitude is calculated via dispersion relations. It is shown
that the integral dispersion relations lead to a better description of the data
for $\sqrt{s}>$5 GeV. Predictions of the considered models for the TOTEM
experiment at LHC energies are given.
\end{abstract}

\section{Introduction}
\label{intro}
High-energy hadron interactions in a soft kinematical region (low or zero transferred
momenta) were and still are an important area of interest for experimentalists and
theoreticians. One of the first tasks of all accelerators always is the measurement of
the total and differential cross sections at small scattering angles. The TOTEM
experiment \cite{totem} is running at the LHC to measure first of all the total
$pp$ cross section at energies 7 TeV and 14 TeV and secondly the differential cross
section of $pp$ scattering in quite a large interval of scattering angle.
From the theoretical point of view, soft physics is beyond the reach of the perturbative
methods of QCD.  The most successful theoretical approach for a description of the
various soft hadron processes is the theory of the analytic $S-$matrix and the methods
of
complex angular momentum.  There are many important
results in $S-$matrix theory that have a general character and do not depend on
additional assumptions such as the existence of the Regge poles. Very useful
examples of such results are the dispersion relations (DR) which the amplitude
of hadron scattering
must satisfy. The dispersion relations for the forward hadron-scattering amplitude is a
subject of the special interest because it can be written in a form which relates
measured quantities. Thus the dispersion relations can be used in two ways. If all terms
in the DR can be calculated, then comparing it with the experimental data verifies the
validity of analyticity, which is one of the main postulates of  the theory.
Alternatively, if  for example some parameters are unknown, we can determine them
by requiring the best agreement of DR with the corresponding experimental data.
In what follows we show how the DR for the forward scattering of $pp, \bar pp,
\pi^{\pm}p$ and $K^{\pm}p$ amplitudes must be applied to describe the experimental data
on the total cross sections and the ratios of the real part to the imaginary part of
amplitudes. We
remind the reader of the main postulates and assumptions which are important
in order to derive DR, describe the procedure for correctly using them, show the
resulting description of the data and make predictions for measurements at LHC
energies.

\section{The main properties of the amplitude needed to derive DR}
\label{sec:1}
{\bf Mandelstam variables.} The analytic $S$-matrix theory postulates that the amplitude
of any
hadronic process $ab\to cd$ is an analytic function of
invariant kinematic variables.
\begin{equation}
s=(p_{a}+p_{b})^{2},\quad
t=(p_{a}-p_{c})^{2},\quad
u=(p_{a}-p_{d})^{2},\qquad
s+t+u =m_{a}^{2}+m_{b}^{2}+m_{c}^{2}+m_{d}^{2}.
\end{equation}
 For the  processes under interest $a^{\pm}p\to a^{\pm} p$ where $a=p,\pi,K, p^{-}\equiv
 \bar p$,
\begin{equation}
s+t+u =2(m_{a}^{2}+m_{p}^{2}).
\end{equation}

\noindent
{\bf Crossing symmetry.} Crossing symmetry means that processes $a+p \longrightarrow
a+p$ ($s$-channel), $a+\bar a \longrightarrow p+\bar p$ ($t$-channel) and $\bar p+p
\longrightarrow a+\bar a$  ($u$-channel)  are described by the limiting values of  one
 analytic function $A(s,t,u)$  taken in different regions of the variables
$s$, $t$ and $u$.
Because only two of three variables $s$, $t$, $u$ are independent, in what follows we
often write $A(s,t)$ instead of $A(s,t,u)$.

\begin{figure}[ht]
\includegraphics[scale=0.45]{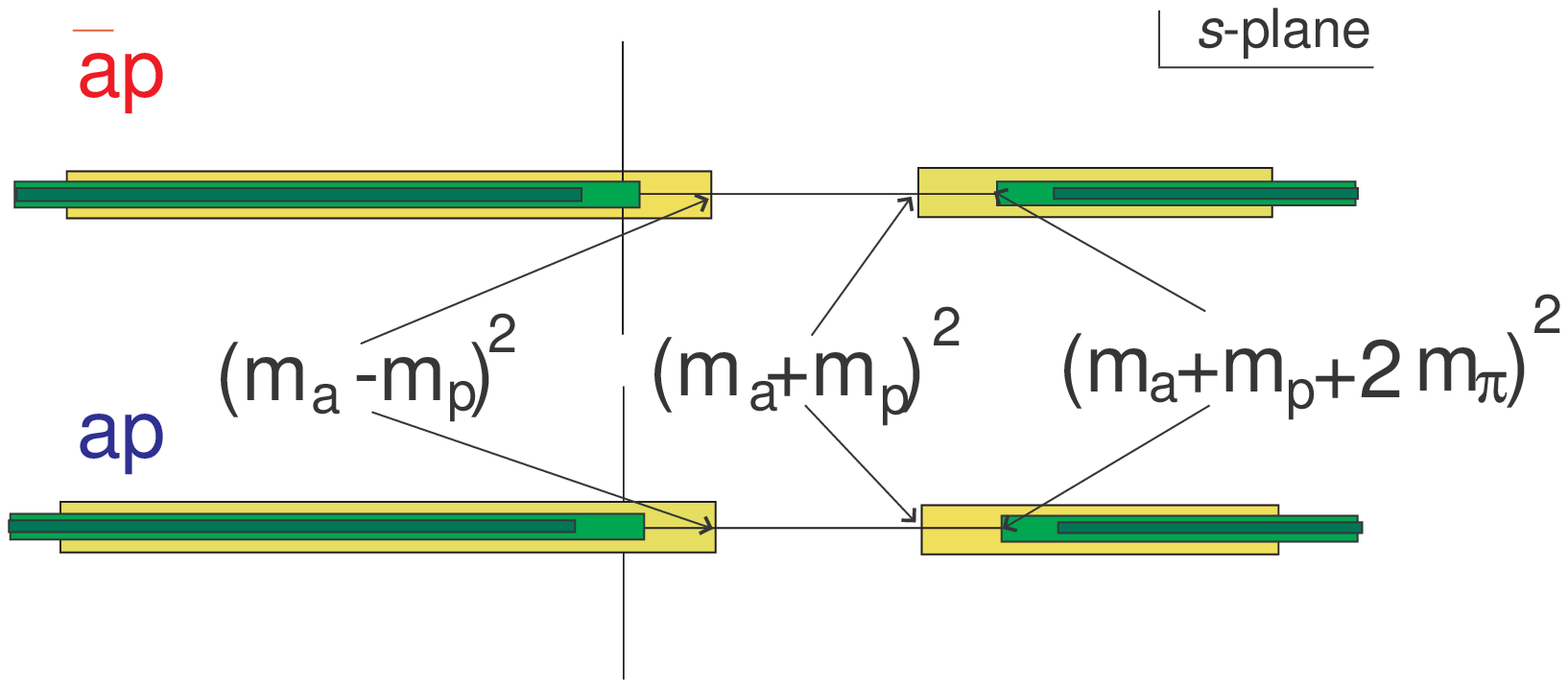}
\hspace{0.8cm}
\includegraphics[scale=0.45]{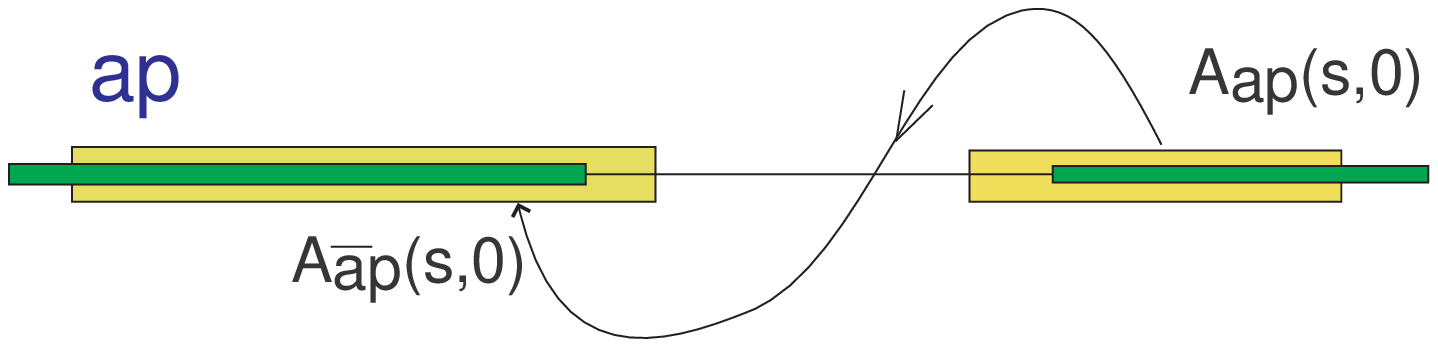}
\caption{\label{fig:singular}Structure of the singularities (left) and analytic
continuation (right) of the
$ap$ and $\bar ap$ elastic-scattering amplitudes}
\end{figure}

\noindent
{\bf Structure of singularities.} The main singularities of $ap$ and $\bar ap$ elastic scattering
amplitudes at $t=0$ are shown in Fig.\ref{fig:singular}. They are: i) the branch points
at $s\geq (m_{a}+m_{p})^{2}$ corresponding to the threshold energies of elastic and
inelastic processes, ii) branch points generated by the thresholds in $u$-channel  at
$s\leq 0$, iii) unphysical branch points (for elastic $\bar ap$ scattering) generated by
$u$-channel states (at $4m_{\pi}^{2}\leq u \leq 4m_{p}^{2}$ ). Thus for amplitude we
have the right-hand and the left-hand cuts shown in Fig. 1 (left).

The physical amplitude of $ap$  elastic scattering is defined at the upper side of the
all cuts from $(m_{a}+m_{p})^{2}$ to $+\infty$, {\it i.e.}
$A_{ap}(s,t)=\lim\limits_{\varepsilon\to 0}A(s+i\varepsilon,t,u )\equiv A_{-}(s,t,u)$ at
$s>(m_{a}+m_{p})^{2}$,  $A_{\bar ap}(u,t)=\lim\limits_{\varepsilon\to 0}A(s,t,u
+i\varepsilon)\equiv A_{+}(s,t,u)$ at $u>(m_{a}+m_{p})^{2}$.  Furthermore, one can
derive from the
definition  that
\begin{equation}\label{eq: pbarp-ampl}
A_{\bar ap}(u,t)=\lim\limits_{\varepsilon\to 0}A(s-i\varepsilon,t,u )\quad
\mbox{at}\quad s+t<0.
\end{equation}
It follows from the above definitions that the amplitude $A_{\bar ap}$ can be obtained
from $A_{ap}$ by analytic continuation as shown on the right-hand part of the Fig.
\ref{fig:singular}.

The amplitudes can have poles at complex values of $s$ corresponding to resonances as well as branch points and corresponding unphysical cuts as for example at $s=4m_{\pi}^{2}$ in $\bar pp$ elastic scattering amplitude. Usually they are considered as small corrections at high energy.

\noindent
{\bf Optical theorem.} For $ap$ and $\bar ap$ it states that
\begin{eqnarray}
\label{eq:stot}
\sigma_{tot}^{\bar ap}(s)\equiv \sigma_{+}=\frac{1}{2m_{p}p}ImA_{\bar ap}(s,0)=\frac{1}
{2q_{s}\sqrt{s}}\, ImA_{+}(s,0),\\
\sigma_{tot}^{ap}(s)\equiv \sigma_{-}=\frac{1}{2m_{p}p}ImA_{ap}(s,0)= \frac{1}
{2q_{s}\sqrt{s}}\, ImA_{-}(s,0)
\end{eqnarray}
where $p$ is the momentum of hadron $a$ in the laboratory system, $q_{s}$ is the
relative momentum of $a$ and $p$ in the center-of-mass system, given by
$q_{s}^{2}=\frac{1}{4s}[s-(m_{a}+m_{p})^{2}][s-(m_{a}-m_{p})^{2}]$ and $A_{\pm}
(s,0)=A^{\bar ap}_{ap}(s,0)$.

\noindent
{\bf Polynomial behaviour.}
It is well known  that the scattering amplitude cannot rise at high $|s|$
faster than a power, i.e. $N$ must exist such that for $|s|\to \infty$ and
$t_{0}<t\leq 0$
\begin{equation}\label{eq: polybound}
|A(s,t)|<|s|^{N}.
\end{equation}

\noindent
{\bf High-energy bounds for cross-sections.}
Total hadron cross sections behave at asymptotic energies in accordance with the well-
known Froissart-Martin-{\L}ukaszuk bound
\begin{equation}\label{eq: FMLb}
\sigma_{t}(s)<\frac{\pi}{m_{\pi}^{2}}\ln^{2} (s/s_{0})\quad  \mbox{at}\quad  s\to
\infty, \quad s_{0}\sim 1 \ \mbox{GeV}^{2}.
\end{equation}
The last inequality means that $|A(s,0)/s^{2}|\to 0\quad  {\rm for}\ |s|\to \infty$.
{\it i.e.}  $N<2$ in Eq.~(\ref{eq: polybound}).

\section{Integral Dispersion Relations (IDR)}
As an analytic function of the variable $s$, the amplitude $A(s,t)$
(in what follows we consider forward scattering amplitude, $t=0$) must satisfy the
dispersion relation which can be derived from Cauchy's theorem for analytic functions:
\begin{equation}\label{eq:Cauchi}
f(z)=\frac{1}{2\pi i}\oint \frac{f(z')}{z'-z}
\end{equation}
where the contour $C$ surrounds the point $z$ and any singularity of
$f(z)$ inside.

Because  hadronic amplitudes $|A(s,0)/s^{2}|\to 0$  while  $|A(s,0)/s|\nrightarrow 0$
at $|s|\to \infty$, it is more convenient to apply Cauchy's theorem to the function
$A(s,0)/((s-s_{0})(s-s_{1}))$ rather than directly to the amplitude. Generally, the
points $s_{0}$ and $s_{1}$ are arbitrary but usually they are chosen at
$s_{0}=s_{1}=2m_{p}^{2}$.
\begin{figure}[ht]
\begin{center}
\includegraphics[scale=0.6]{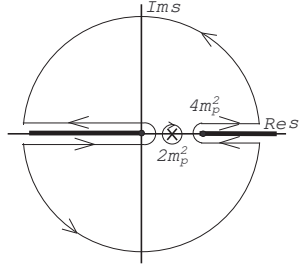}
\caption{Deformation of an integration contour in the Cauchy integral for $A(s,0)$}
\label{fig:contour}
\end{center}
\end{figure}

Deforming the integration contour $C$ as shown in Fig. \ref{fig:contour} in Cauchy's
theorem (more details can be found in the books \cite{Collins,BP,DLND}),  and taking
the circle to an infinite radius, so that its integral tends to $0$
(because  $|A(s,0)/s^{2}|\to 0$ at $|s|\to \infty$), one can write
\begin{eqnarray}\label{eq:dispers III}
A(s,0)=&A(s_{0},0)+(s-s_{0})A'(s_{0},0)+
\frac{(s-s_{0})^{2}}{2\pi}\bigg [\int\limits_{s_{tr}}^{\infty}
\frac{D_{s}(s',0)}{(s'-s_{0})^{2}(s'-s)}ds'
+\int\limits_{-\infty}^{\tilde s_{tr}}
\frac{D_{s}(s',0)}{(s'-s_{0})^{2}(s'-s)}ds'\bigg  ]\nonumber\\
=&A(s_{0},0)+(s-s_{0})A'(s_{0},0)+
\frac{(s-s_{0})^{2}}{2\pi}\bigg [\int\limits_{s_{tr}}^{\infty}
\frac{D_{s}(s',0)}{(s'-s_{0})^{2}(s'-s)}ds'
+\int\limits_{s_{tr}}^{\infty}
\frac{D_{u}(u',0)}{(u'-u_{0})^{2}(u'-u)}du'\bigg ]
\end{eqnarray}
where $s_{tr}=(m_{a}+m_{p})^{2},  \tilde s_{tr}=(m_{a}-m_{p})^{2}$ for the amplitudes
$A_{ap}^{\bar ap}(s,0)$  and
\begin{equation}\label{eq: discontin}
D_{s}(s,t)=\frac{1}{i}[A(s+i\epsilon,t,u)-A(s-i\epsilon,t,u)],\qquad
D_{u}(u,t)=\frac{1}{i}[A(s,t,u+i\epsilon)-A(s,t,u-i\epsilon)]
\end{equation}
are the discontinuities of the amplitude accross the cuts.

After some simple transformations one can obtain the  standard  form of the integral dispersion relations written  in the laboratory system  ($s=2m_{p}(E+m_{p}), u=2m_{p}(-E+m_{p})$, the point $s_{0}=u_0$ corresponding to $E_{0}=0$ ):
\begin{equation}\label{eq:2s-idr}
\rho_{\pm}\sigma_{\pm}=\frac{A_{\pm}(s_{0},0)}{2m_{p}p}+\frac{E\,
A_{\pm}'(s_{0},0)}{p} +\frac{E^{2}}{\pi p}{\rm
P}\int\limits_{m_{a}}^{\infty}\left
[\frac{\sigma_{\pm}}{E'^{2}(E'-E)}+\frac{\sigma_{\mp}}{E'^{2}(E'+E)}\right
]p'\, dE'
\end{equation}
where $\rho_{\pm}=ReA_{\pm}(s,0)/ImA_{\pm}(s,0)$, $A'(z,0)=dA(z,0)/dz$,
$s_{0}=(m_{a}+m_{p})^{2}$ and
\begin{equation}\label{eq:x-sym}
A_{+}(s_{0})=A_{-}(s_{0}), \qquad A'_{+}(s_{0})=-A'_{-}(s_{0}).
\end{equation}

We would like to note that in reviews on high-energy physics \cite{BlockCahn} and \cite{Block}  the dispersion relation with two subtractions contains missprints. Namely, in \cite{BlockCahn}  the indexes ``$pp$'' and ``$\bar pp$'' must be rearranged in the integrands of the Eqs. (4.85) and (4.86). In the Eqs. (203) and (204) the sign ``-'' in the integrand must be replaced for the ``+''.

It was confirmed by the COMPETE analysis \cite{COMPETE}, as well as in \cite{BlockKang},
that there are no indications of an odderon contribution to $\sigma_{tot}(s)$ and $\rho(s)$.
This means that $\Delta \sigma_{t}(s)=\sigma_{t}^{(\bar ap)}(s)-\sigma_{t}^{(ap)}(s)\to 0$
at $s\to \infty$. If this is so, one should not apply the IDR in the form of Eq.
(\ref{eq:2s-idr}) with  free unknown constants
$A(s_{0}),  A'(s_{0})$ even if $\sigma_{t}^{(a^{\pm} p)}(s)\propto \ln^{2}s$, i.e. $|A(s,0)|
\propto s\ln^{2}s$ .

Indeed, let us consider the relation (\ref{eq:2s-idr}), insert $E$ in the integrand and
write it as $(E-E')+E'$ in the first term and $(E+E')-E'$
in the second one. After simple transformations we obtain
\begin{equation}\label{eq:dis2sub}
\begin{array}{lll}
\rho_{\pm}\, \sigma_{\pm} & = & \dst \frac{1}{p}\left \{
A_{\pm}(s_{0},0)/2m_{p}+E\, \left [ A_{\pm}'(s_{0},0)\mp
\frac{1}{\pi}\int\limits_{m_{a}}^{\infty}\frac{\sigma_{+}-\sigma_{-}}{E'^{2}}p'\,
dE'\right ]\right \}\\
 & + & \dst \frac{E}{\pi p}{\rm P}\int\limits_{m_{a}}^{\infty}\left
[\frac{\sigma_{\pm}}{E'(E'-E)}-\frac{\sigma_{\mp}}{E'(E'+E)}\right ]p'\,
dE'.
\end{array}
\end{equation}

Now one can show that if $\Delta \sigma =\sigma_{+}-\sigma_{-}\to 0$ then
\begin{equation}\label{eq:delsig}
 A_{\pm}'(s_{0},0)=\pm
\frac{1}{\pi}\int\limits_{m_{a}}^{\infty}\frac{\sigma_{+}-\sigma_{-}}{E'^{2}}p'\,dE'.
\end{equation}
This can be proven taking into account that $\sigma_{+}(s')-\sigma_{-}(s')=(-i/4m_{p}p')
[A^{(-)}(s'+i\varepsilon)-A^{(-)}(s'-i\varepsilon)]$, where $A^{(-)}(s)$ is the crossing-odd
part of the amplitudes $A_{\pm}(s)$
\begin{equation}
A_{+}(s)\pm A_{-}(s)=2A^{(\pm)}(s).
\end{equation}

Thus, if the odderon does not contribute to the amplitudes $A_{ap}^{\bar ap}(s,0)$ then  the
constants $A'_{\pm}(s_{0},0)$ cannot be considered as free parameters, and their values can
be
calculated explicitly. Furthermore the IDR of the Eq.(\ref{eq:2s-idr}) is reduced to
the form
\begin{equation}\label{eq:1s-idr}
\rho_{\pm}\, \sigma_{\pm} =\frac{A_{\pm}(s_{0},0)} {2m_{p}p} + \frac{E}{\pi p}{\rm
P}\int\limits_{m_{p}}^{\infty}\left
[\frac{\sigma_{\pm}}{E'(E'-E)}-\frac{\sigma_{\mp}}{E'(E'+E)}\right ]p'\,dE'
\end{equation}
which is the IDR with one subtraction. Dispersion relations of this form  were first
applied to data analysis by P. S\"{o}ding \cite{Soding} in 1964. It was then believed in
that cross sections go to a constant at asymptotic energies. Therefore an application of the
IDR with one subtraction was completely justified. But now we know that cross sections are
rising with energy.  Moreover, if we want to check the odderon hypothesis within the IDR
method, we must use the relation (\ref{eq:2s-idr}) at least for $pp$ and $\bar pp$ cross
sections and $\rho$ ratios. Note that because of its negative P-parity, the odderon does not
contribute to meson-nucleon amplitudes. Hence for $\pi p$ and $Kp$ amplitudes we have to use
the IDR (\ref{eq:1s-idr}) with one substraction.

\section{Phenomenological application of the IDR for meson-proton and proton-proton forward-
scattering amplitudes}

We consider three explicit models for high-energy $p^{\pm} p, \pi^{\pm} p, K^{\pm}p$ total
cross sections and corresponding $\rho$ ratios calculated through integral dispersion
relations. We compare two possibilities for $p^{\pm} p$ cross sections, with and without an
odderon contribution. In the first case, the IDR with two subtractions, Eq.(\ref{eq:2s-idr}),
is used to calculate $\rho_{pp}$ and $\rho_{\bar pp}$, while in the second case the
IDR  with one subtraction, Eq.(\ref{eq:1s-idr}), is applied to calculate all $\rho$ ratios.

\subsection{\bf Low-energy part of the dispersion integral.}\label{sec:low-energy}
\begin{figure}[ht]
\begin{center}
 \includegraphics[width=0.9\textwidth]{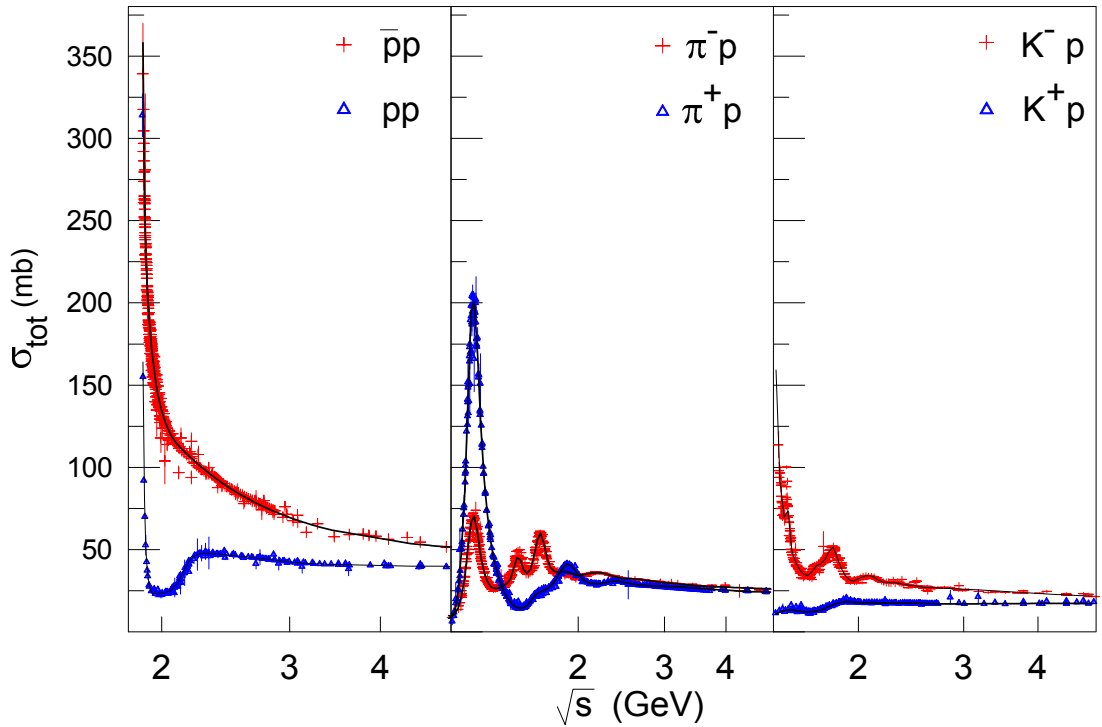}
\end{center}
\caption{Description of the low-energy cross sections.}
\label{fig:3}
\end{figure}

A high-energy parametrizations for the
total cross sections based on the three pomeron models describe the data (all data used for presented analysis are taken from the standard set of the Particle Data Group \cite{PDG}) at $\sqrt{s}\geq
\sqrt{s_{min}}$= 5 GeV well. However, in order to calculate $\rho (s)$, we have to
integrate the total cross sections from the threshold up to infinity. Thus we need an
analytic form for the total cross sections at low energy. Therefore, we parametrize the
cross sections for each process at low energies by some function which can have any number
of parameters. The main aim is to describe the data as well as possible. Then when the high-energy data are fitted to in the various pomeron models, all these low-energy parameters are
fixed. We only slightly change the low-energy parameterization to ensure continuity of the
cross sections at the point $s=s_{min}$. Thus, in the low-energy parametrization we keep one
free parameter for each cross section. The details of the parametrization of low-energy
cross section play an auxiliary role and do not influence the results at $\sqrt{s}\geq
\sqrt{s_{min}}$.
The quality of the data description is quite good as can be seen from Fig.\ref{fig:3}.

\begin{figure}[ht]
\begin{center}
  \includegraphics[width=0.45\textwidth]{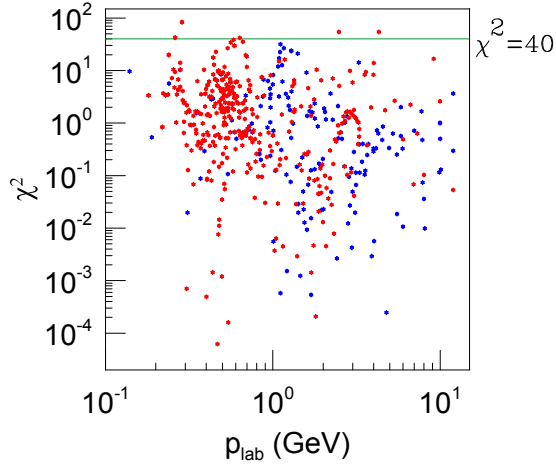}
\end{center}
\caption{$\chi^{2}$ in the fit of the low-energy $pp$ and $\bar pp$ cross sections.}
\label{fig:4}
\end{figure}

However we would like to comment on the obtained $\chi^{2}/N_{p}$ ($\approx$ 3.9  for the
whole set of data (number of points $N_{p}$=1932). The data are strongly spread around the
main group of points, as it is seen from the Fig.\ref{fig:4}. There are a few points
deviating quite far from them, and some of these points  (for example in the $pp$ set only 6
points) individually contribute more than 40 to $\chi^{2}$ . If we exclude their
contribution we obtain a reasonable value $\chi^{2}/N_{p} \approx 1.5$. In our opinion this
quality of data description is  acceptable in order to have a sufficiently precise value of
the dispersion integral from the threshold to $\sqrt{s}$=5 GeV.

\subsection{\bf High-energy pomeron models.}
We consider three models leading to different asymptotic behavior for the
total cross sections. We start from the explicit parameterization of the
total $pp$ and $\bar pp$ cross sections, then, to find the ratios of the
real part to the imaginary part, we apply the IDR making use of the above-described
parameterizations  and  calculating  the low-energy
part of the dispersion integral.

All the models include the contributions of the pomeron ${\cal P}$, of crossing-even and
cros\-sing-odd
reggeons {\cal R} (we consider these two reggeons as effective ones to avoid increasing the
number of parameters as would be the case if we included the full set
of  secondary reggeons, $f, \omega, \rho, a_{2}$) and of the odderon (for $pp$ and $\bar
pp$).

\begin{equation}\label{eq:sigmod}
Im A_{ap}^{\bar ap}(s,0)= {\cal P}(z)+{\cal R}_{+}(z)\pm {\cal R}_{-}(z)\pm {\cal O}(z),
\end{equation}
\begin{equation}\label{eq:reggeon}
{\cal R}_{\pm}(z)=g_{\pm}z^{\alpha_{\pm}(0)},
\end{equation}
\begin{equation}\label{eq:zt}
z=|\cos\vartheta _{t}|=\frac{t+2(s-m_{p}^{2}-m_{a}^{2})}{\sqrt{(4m_{p}^{2}-t)(4m_{a}^{2}-t)}}=\frac{t+4Em_{p}^{2}}{\sqrt{(4m_{p}^{2}-t)(4m_{a}^{2}-t)}}
\end{equation}
where $\vartheta _{t}$ is the scattering angle in the cross channel. If $t=0$ then $z=E/m_{a}$.

\noindent{\bf Simple-pole-pomeron model (SP).} In this model, the intercept of the
pomeron is larger than unity. In contrast with well-known Donnachie-Landshoff model  \cite{D-L}, we add in the amplitude a simple pole (with $\alpha_{\cal P}(0)=1$)
contribution
\begin{equation}\label{eq:SPim}
{\cal P}(E)=g_{0}z+g_{1}z^{\alpha_{\cal P}(0)}.
\end{equation}
 In this model, we write the odderon contribution in the form
\begin{equation}\label{eq:Sodd}
{\cal O}(E)=g_{od}z^{{\alpha_{\cal O}(0)}}, \quad \alpha_{\cal O}(0)\leq \alpha_{\cal P}(0).
\end{equation}

\noindent{\bf Dipole-pomeron model (DP).} The pomeron in this model is a double
pole in the complex-angular-momentum plane with intercept $\alpha_{{\cal
P}}(0)=1$.
\begin{equation}\label{eq:DPim}
{\cal P}(E)=g_{0}z+g_{1}z \ln z,
\end{equation}
\begin{equation}\label{eq:Dodd}
{\cal O}(E)=g_{od}z.
\end{equation}

\noindent{\bf Tripole-pomeron model (TP)}. This pomeron is the hardest complex
$j$-plane singularity allowed by unitarity, it is a pair of branch points which collide when
$t\to 0$ and produce a triple pole  at  $j=1$
\begin{equation}\label{eq:TPim}
{\cal P}(E)=g_{0}z+g_{1}z\ln z+g_{2}z\ln^{2}z.
\end{equation}
\begin{equation}\label{eq:Todd}
{\cal O}(E)=g_{1od}z+g_{2od}z\ln z.
\end{equation}

The real part of amplitude can be calculated in two ways. It is obtained either by IDR (for
$pp$ and $\bar pp$, with Eq. (\ref{eq:2s-idr}) if the odderon is taken into account and with
Eq.(\ref{eq:1s-idr}) if not, or by the derivative-dispersion-relation method (alternatively
one can use explicit parameterizations  for both the imaginary part and the real part of the
amplitude). Here we present the results for the IDR method. The second method is discussed
and used in \cite{CMS-1,CMS-2,AvilaMenon}.

We can compare the fits using IDR with those based on standard asymptotic expressions
for the amplitudes. These are built as follows.
The contribution of Regge poles of signature $\tau$ ( +1  or -1) to the amplitude is
\begin{equation}\label{eq:regge}
A_{R_{\tau}}(s,0)=\eta_{\tau}(\alpha_{R}(0))g_{R}z_{t}^{\alpha_{R}(0)}
\end{equation}
where $\eta_{\tau}(\alpha_{R})$ is the signature factor
\begin{equation}\label{eq:signat}
\dst \eta_{\tau}(\alpha_{R})=\frac{1+\tau \exp(-i\pi \alpha_{R})}{-\sin(\pi \alpha_{R})}
=\left \{
\begin{array}{l}
-\exp(-i\pi \alpha_{R}/2)/\sin(\pi \alpha_{R}/2), \quad \tau=+1,\\
-i\exp(-i\pi \alpha_{R}/2)/\cos(\pi \alpha_{R}/2), \quad \tau=-1.
\end{array}
\right .
\end{equation}
The pomeron, odderon and reggeon contributions to the $pp,\ \bar pp$ scattering amplitudes due to the form  (\ref{eq:signat}) of the signature factor can be written as follows
\begin{equation}\label{eq:-is}
\begin{array}{lll}
A^{\bar ap}_{ap}(s,0)&=&-{\cal P}(-i\tilde s)-{\tilde R}_{+}(-i\tilde s)\mp
i{\tilde R}_{-}(-i\tilde s)\mp i{\tilde {\cal O}}_{-}(-i\tilde s)
\end{array}
\end{equation}
where $\tilde s=s/s_{0}, s_{0}=1$ GeV$^{2}$ and
\begin{equation}\label{eq:rresidue}
\tilde R_{\pm}(-i\tilde s)=\left \{
\begin{array}{l}
{\cal R}_{+}(-i\tilde s)/\sin(\pi \alpha_{+}/2), \quad \tau=+1,\\
{\cal R}_{-}(-i\tilde s)/\cos(\pi \alpha_{-}/2), \quad \tau=-1.
\end{array}
 \right .
\end{equation}
The odderon contribution ${\cal O}$ has a similar form if it is chosen as a simple pole.
An advantage of the presentation (\ref{eq:-is}) is that the cross sections in the models
(\ref{eq:reggeon}) and (\ref{eq:rresidue}) have the same form. If the asymptotic
normalization $\sigma_{t}(s)=ImA(s,0)/s) $ is chosen then Eq. (\ref{eq:-is}) is a standard
analytic  parametrization  in its asymptotic  form.
We denote a fit with such expressions for the amplitudes as a ``$-is$ fit".

We present here the results of the fit using IDR  with the standard optical theorem
(\ref{eq:stot}) and briefly compare it with ``$-is$'' fits.

\section{Fit results.}
\subsection{\bf The experimental data and the fitting procedure.}
We apply the dispersion relation  method to the description and analysis of experimental data not only for
$pp$ and $\bar pp$ total cross sections and ratios of the real part to the imaginary part of
the forward scattering amplitudes. This has been done in the number of papers \cite{Block,CMS-1,CMS-2,CMSL,AvilaMenon,Ishida-Igi,AM} (see also refs in these papers).
We consider here additionally $\pi^{\pm} p$ and $K^{\pm} p$ data. All the data are taken
from the standard set \cite{PDG}. There are 411 points for total cross sections
$\sigma_{t}^{a^{\pm}p}(s)$ and 131 points for ratios $\rho^{a^{\pm}p}(s)$ at $\sqrt{s}\geq $
5 GeV.

\subsection{\bf The odderon contribution.}
\begin{figure}[ht]
\begin{center}
\includegraphics[width=0.6\textwidth]{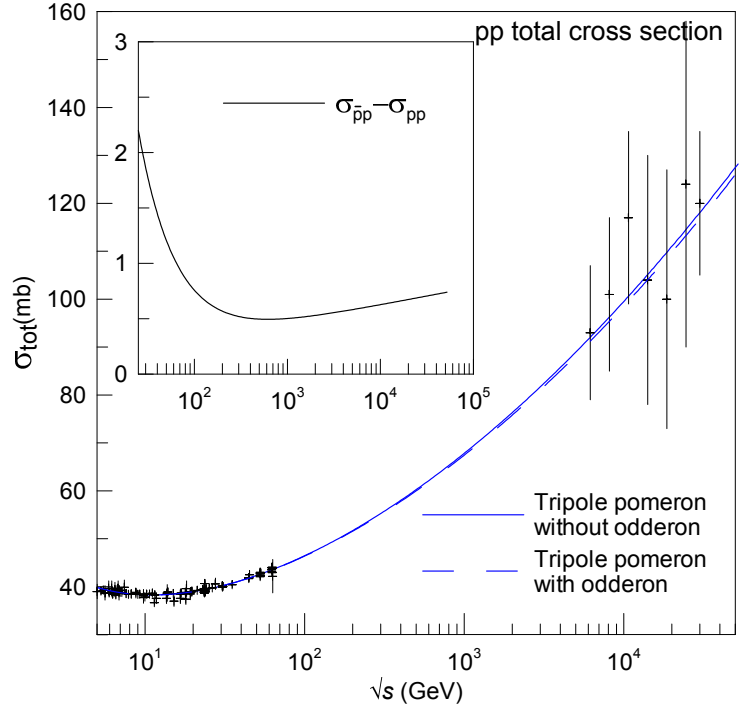}
\end{center}
\caption{Tripole pomeron model with and without Odderon contribution.}
\label{fig:sig-odd}
\end{figure}

\begin{figure}[ht]
\begin{center}
\includegraphics[width=0.5\textwidth]{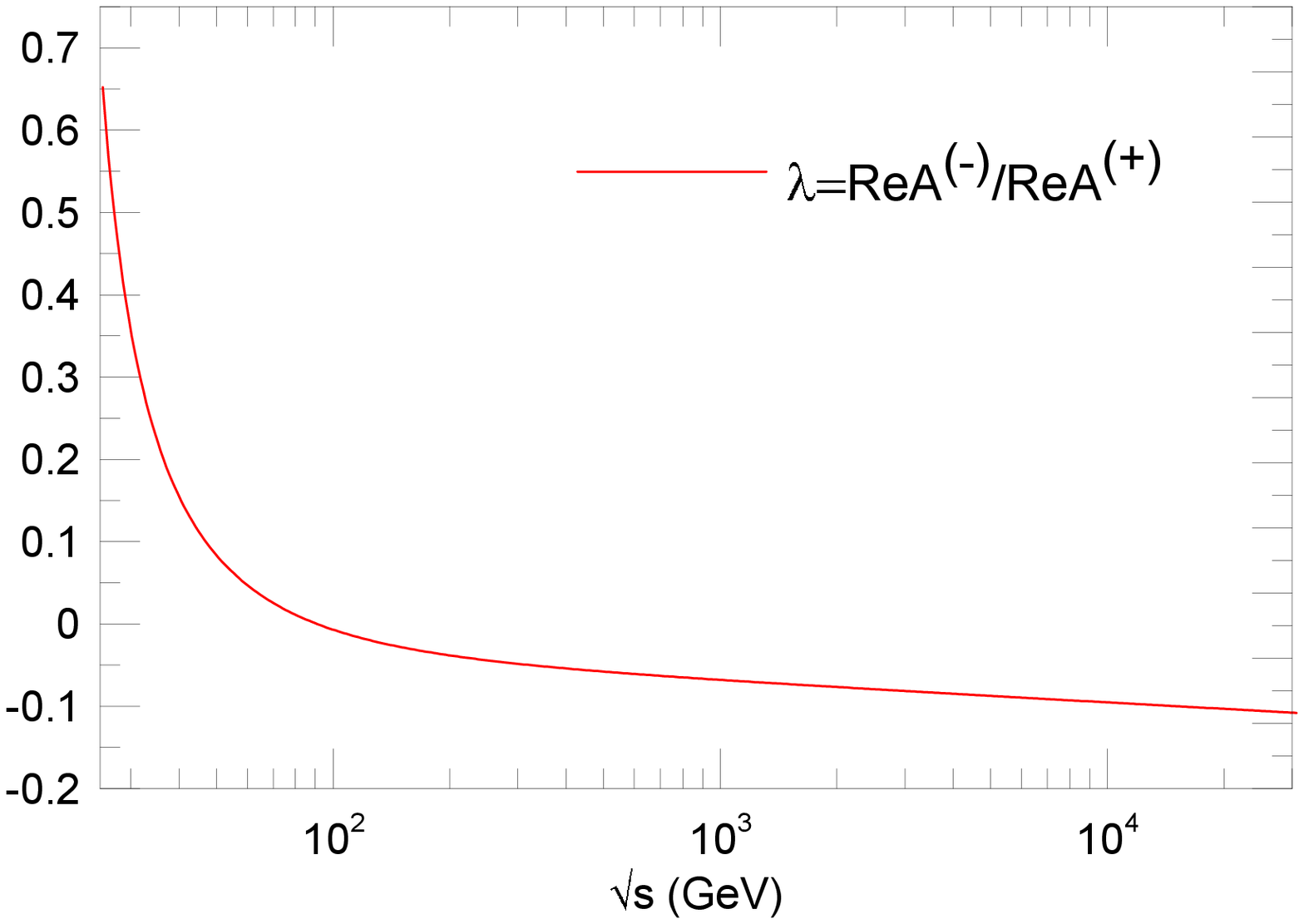}
\end{center}
\caption{The odderon contribution to the real part of the amplitude in the tripole pomeron model at high energy.}
\label{fig:rat-odd}
\end{figure}

The odderon terms defined in Eqs.(\ref{eq:Sodd},\ref{eq:Dodd},\ref{eq:Todd}) in the
corresponding pomeron models give small contribution to the $pp$ and $\bar pp$ cross
sections. At LHC energies, the Reggeon contributions ${\cal R}_{\pm}$ are negligible,
therefore $\Delta \sigma_{t}=\sigma_{t}^{\bar pp}-\sigma_{t}^{pp}$ is dominated by
the odderon contribution. The fit to the data shows that the odderon contribution to the
imaginary part of the forward scattering $pp$ and $\bar pp$ amplitude is very small.
However, the real part of the odderon contribution is about 10\% of the real part
of the pomeron contribution, which can calculated at high energy as follows
\begin{equation}
\lambda=\frac{\rho_{\bar pp}\sigma_{t}^{\bar pp}-\rho_{pp}\sigma_{t}^{pp}}{\rho_{\bar
pp}\sigma_{t}^{\bar pp}+\rho_{pp}\sigma_{t}^{pp}}=\frac{ReA^{(-)}} {ReA^{(+)}}.
\end{equation}
The behavior  of $\Delta \sigma$ and of $\lambda$ at energies $\sqrt{s}>$25 GeV are shown in
Fig. \ref{fig:sig-odd} and Fig. \ref{fig:rat-odd}. One can barely see a very small odderon
contribution to the total cross section: ($\Delta \sigma/\sigma \leq 0.5\%$. Indeed, in the
tripole-pomeron model $\sigma \propto \ln^{2}s$ and $\Delta
\sigma \propto \ln s$). Nevertheless, Fig. \ref{fig:rat-odd}  shows that the contribution of the
odderon to the real part of the amplitudes in the TeV-energy region is sizeable (about
10\%). The tripole-pomeron
model is shown as an example, as the dipole- and simple-pole pomeron models show the same
odderon effects.

However, comparing the values of the $\chi^{2}$ obtained in the considered models, one can
conclude that the odderon contribution is not significant for data at $t=0$.
The values of the $\chi^{2}/dof$ obtained in the considered models are given in the Table
\ref{tab:chi2}.

\begin{table}
\caption{The values of $\chi^{2}/dof $ in three pomeron models with and without odderon
contribution}
\label{tab:chi2}
\begin{tabular}{lccc}
\hline\noalign{\smallskip}
&  \multicolumn{3}{c}{$\chi^{2}/dof$}\\
\cline{2-4}  
 & Simple pole pomeron & Dipole pomeron & Tripole pomeron  \\
\noalign{\smallskip}\hline\noalign{\smallskip}
With odderon & 0.974 & 0.974 & 0.960 \\
No odderon & 0.976 & 0.974 & 0.963 \\
\noalign{\smallskip}\hline
\end{tabular}
\end{table}
\subsection{\bf Models without an odderon.}
\begin{figure}[ht]
\begin{minipage}{7.6cm}
  \includegraphics[width=1.0\textwidth]{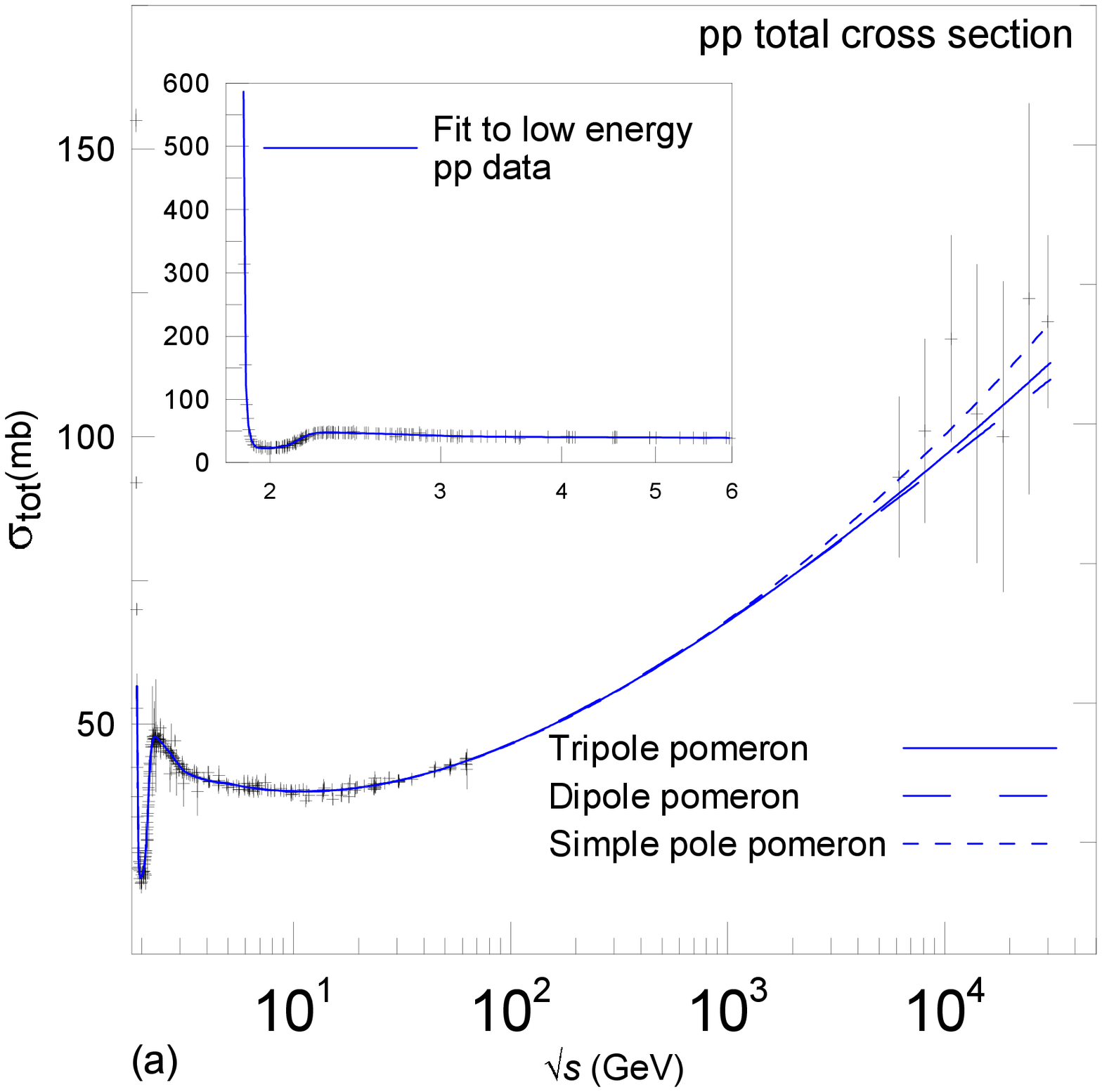}
 \end{minipage}
\begin{minipage}{7.6cm}
  \includegraphics[width=1.0\textwidth]{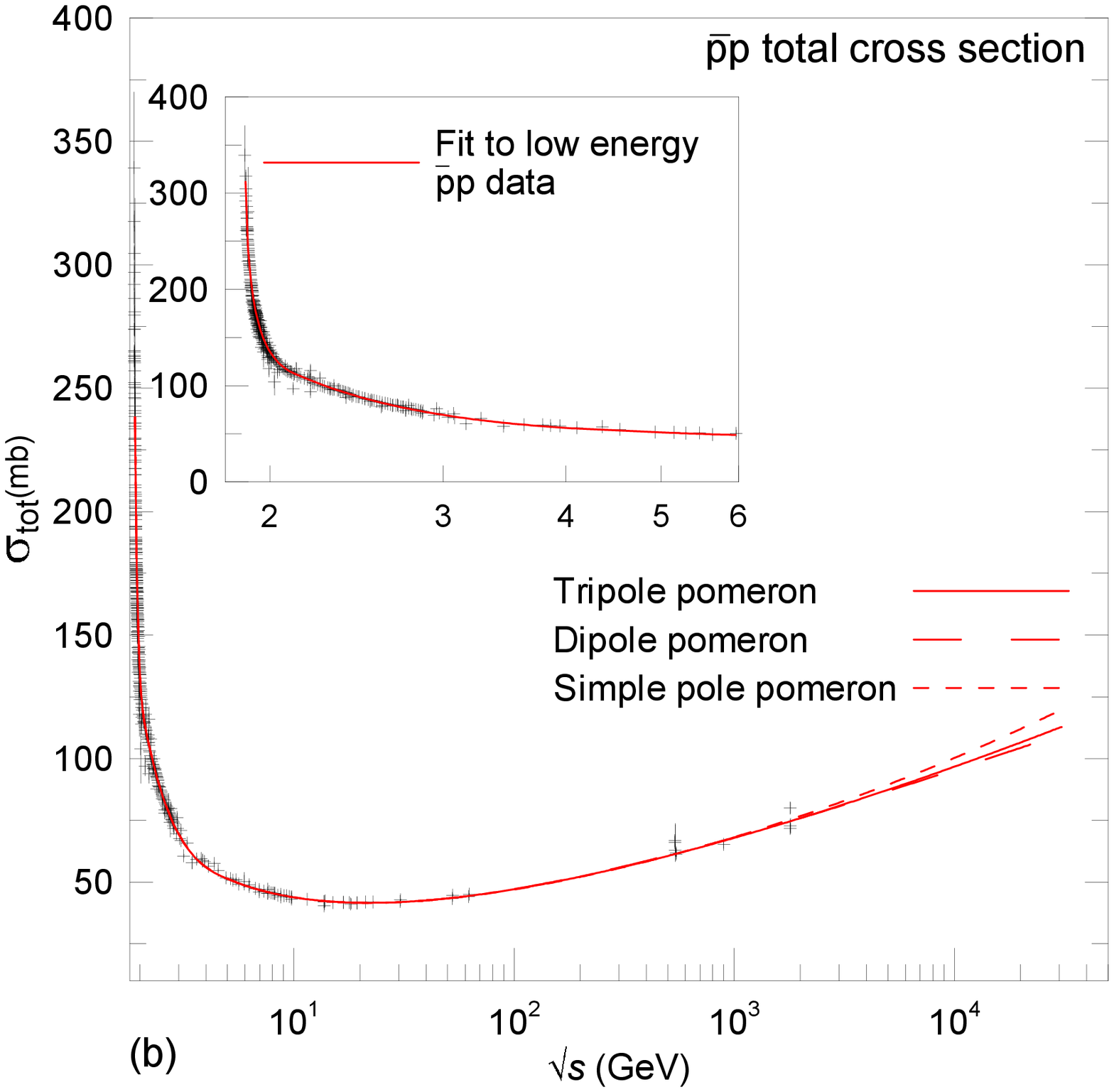}
 \end{minipage}
\caption{Total $pp$ (a) and $\bar pp$ (b) cross sections}
\label{fig:sig-pp-pap}
\end{figure}

\begin{figure}[ht]
\begin{minipage}{7.6cm}
  \includegraphics[width=1.0\textwidth]{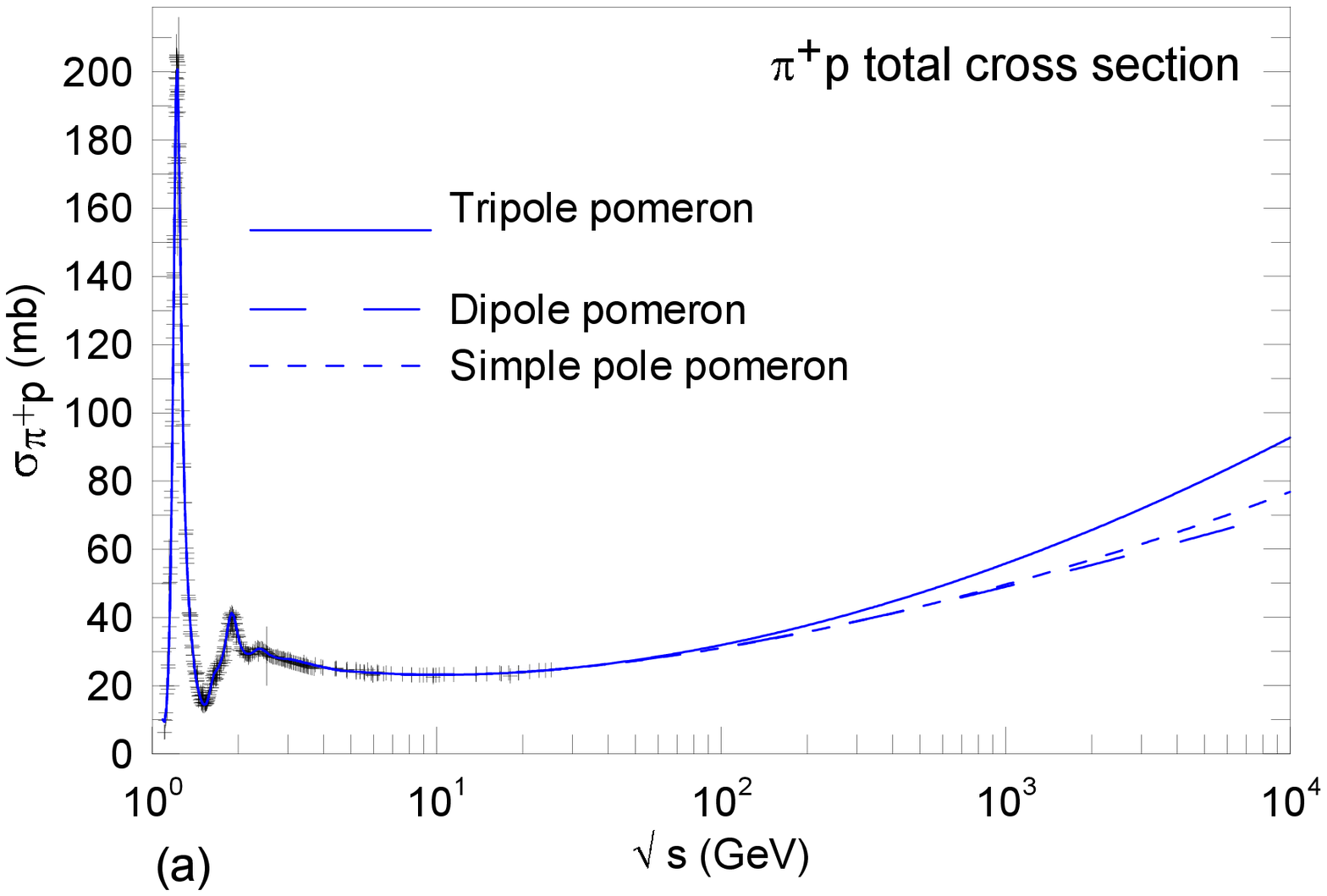}
 \end{minipage}
\begin{minipage}{7.6cm}
  \includegraphics[width=1.0\textwidth]{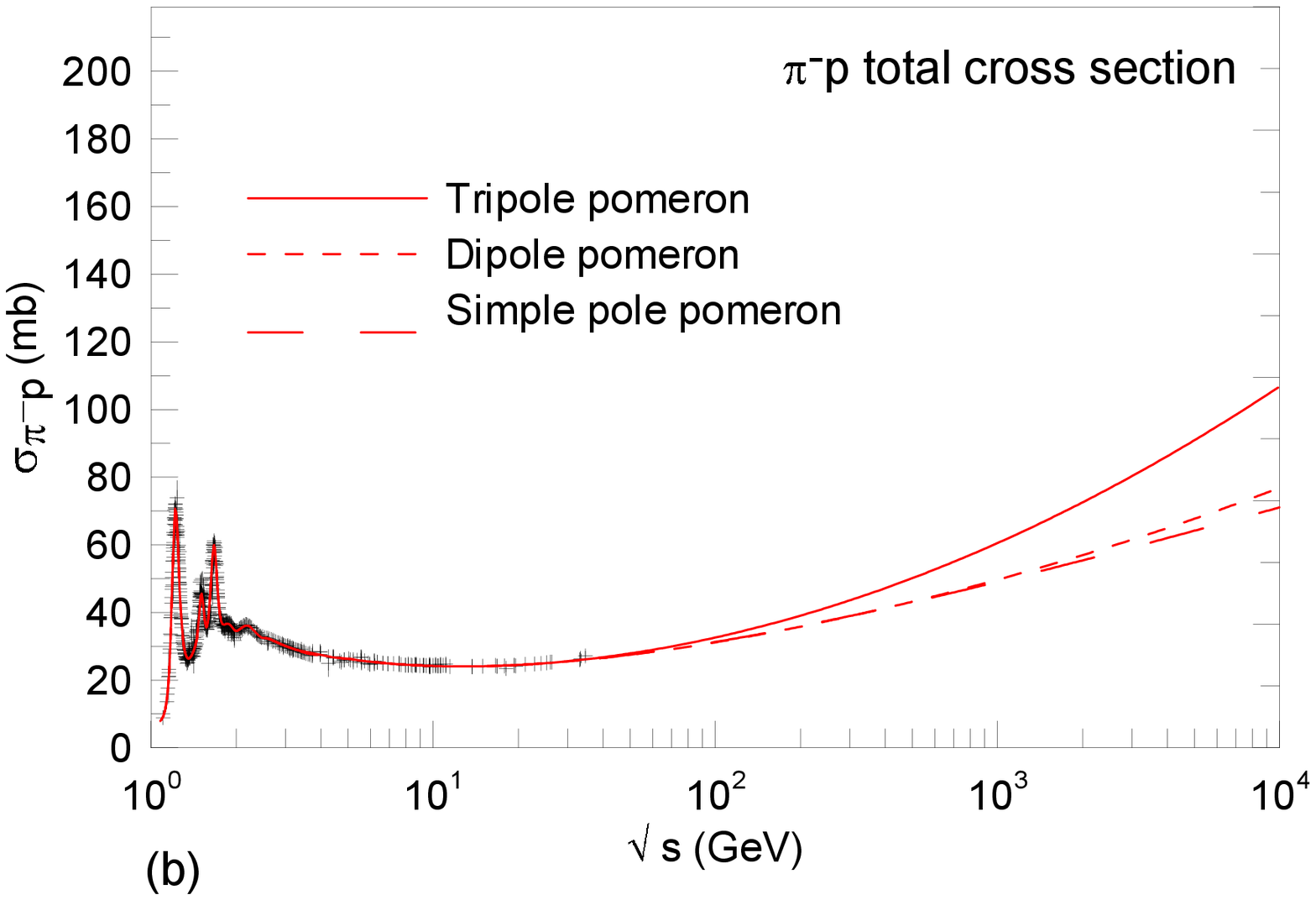}
 \end{minipage}
\caption{Total $\pi^{+}p$ (a) and $\pi^{-}p$ (b) cross sections}
\label{fig:sig-pip-pim}
\end{figure}

\begin{figure}[ht]
\begin{minipage}{7.6cm}
  \includegraphics[width=1.0\textwidth]{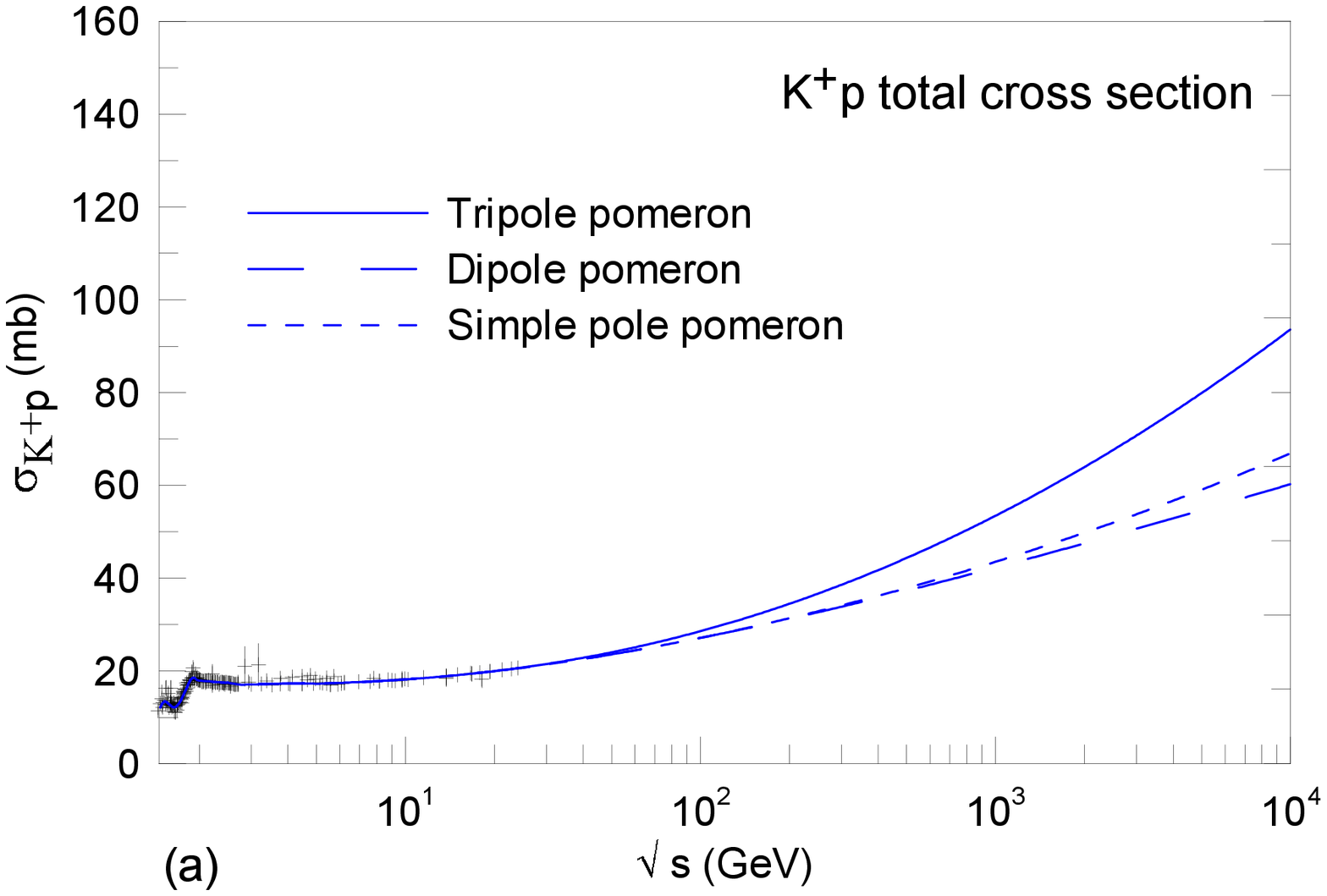}
 \end{minipage}
\begin{minipage}{7.6cm}
  \includegraphics[width=1.0\textwidth]{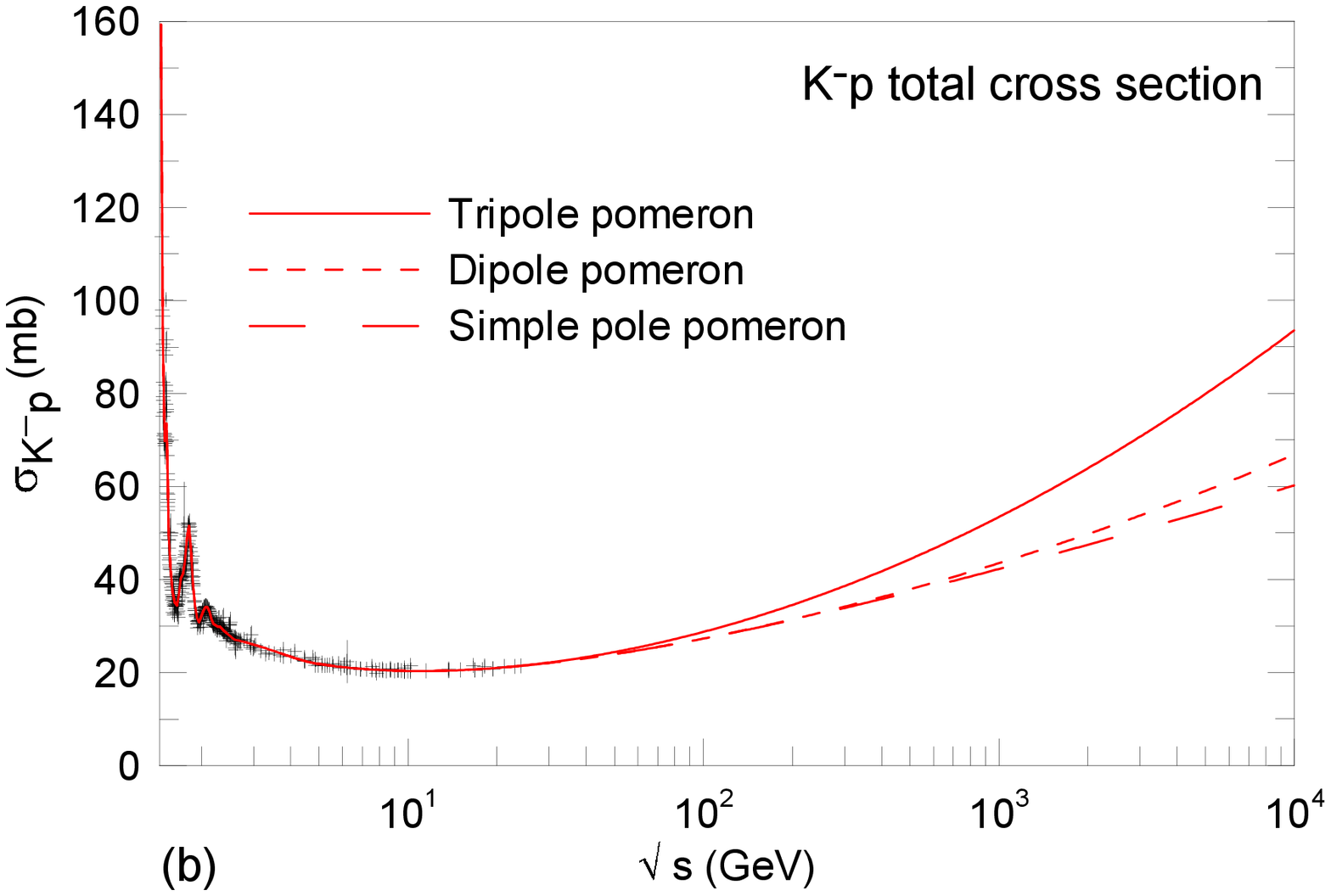}
 \end{minipage}
\caption{Total  $K^{+}p$ (a) and $K^{-}p$ (b) cross sections}
\label{fig:sig-kp-km}
\end{figure}

\begin{figure}[ht]
\begin{minipage}{7.6cm}
  \includegraphics[width=1.0\textwidth]{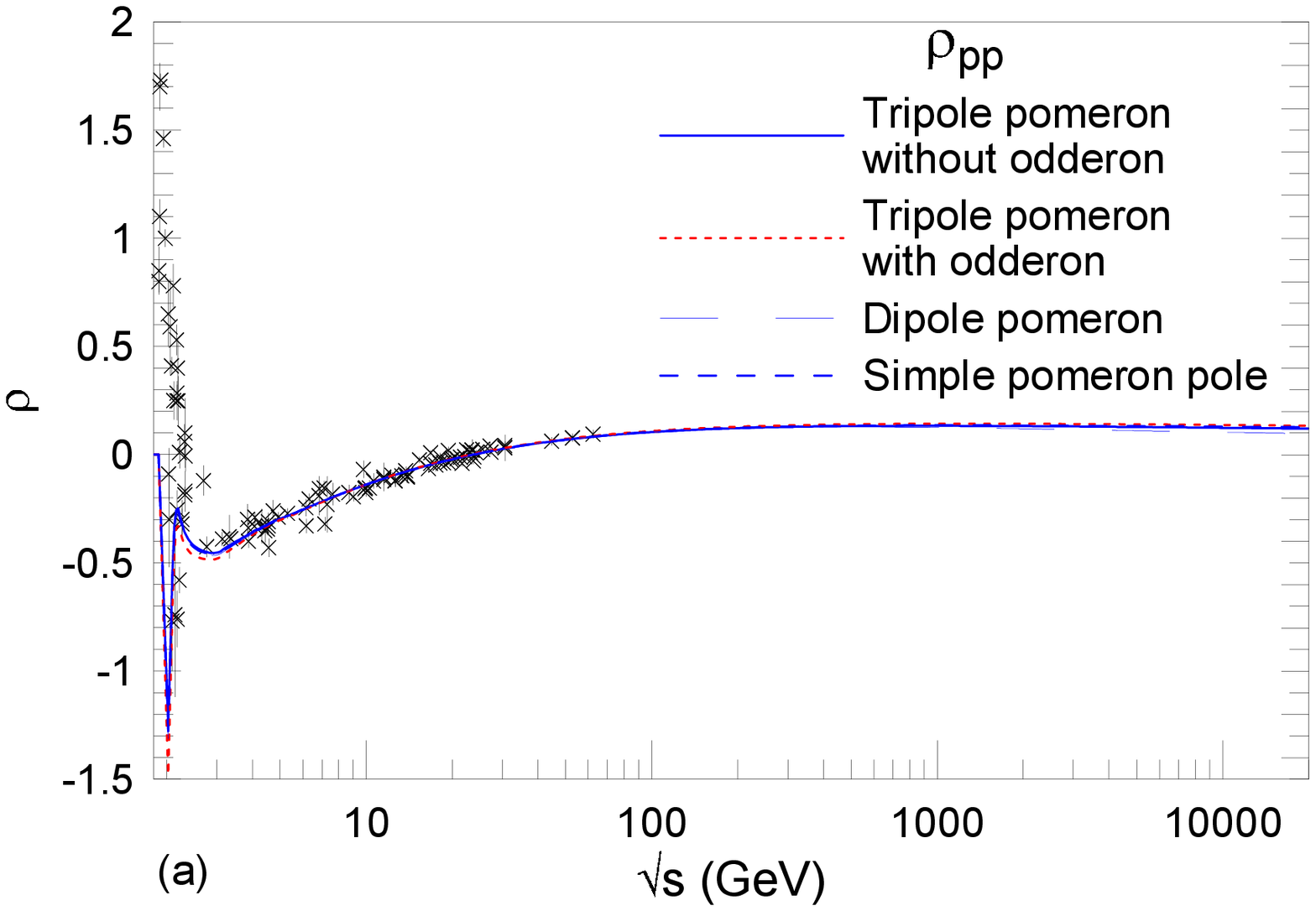}
 \end{minipage}
\begin{minipage}{7.6cm}
  \includegraphics[width=1.0\textwidth]{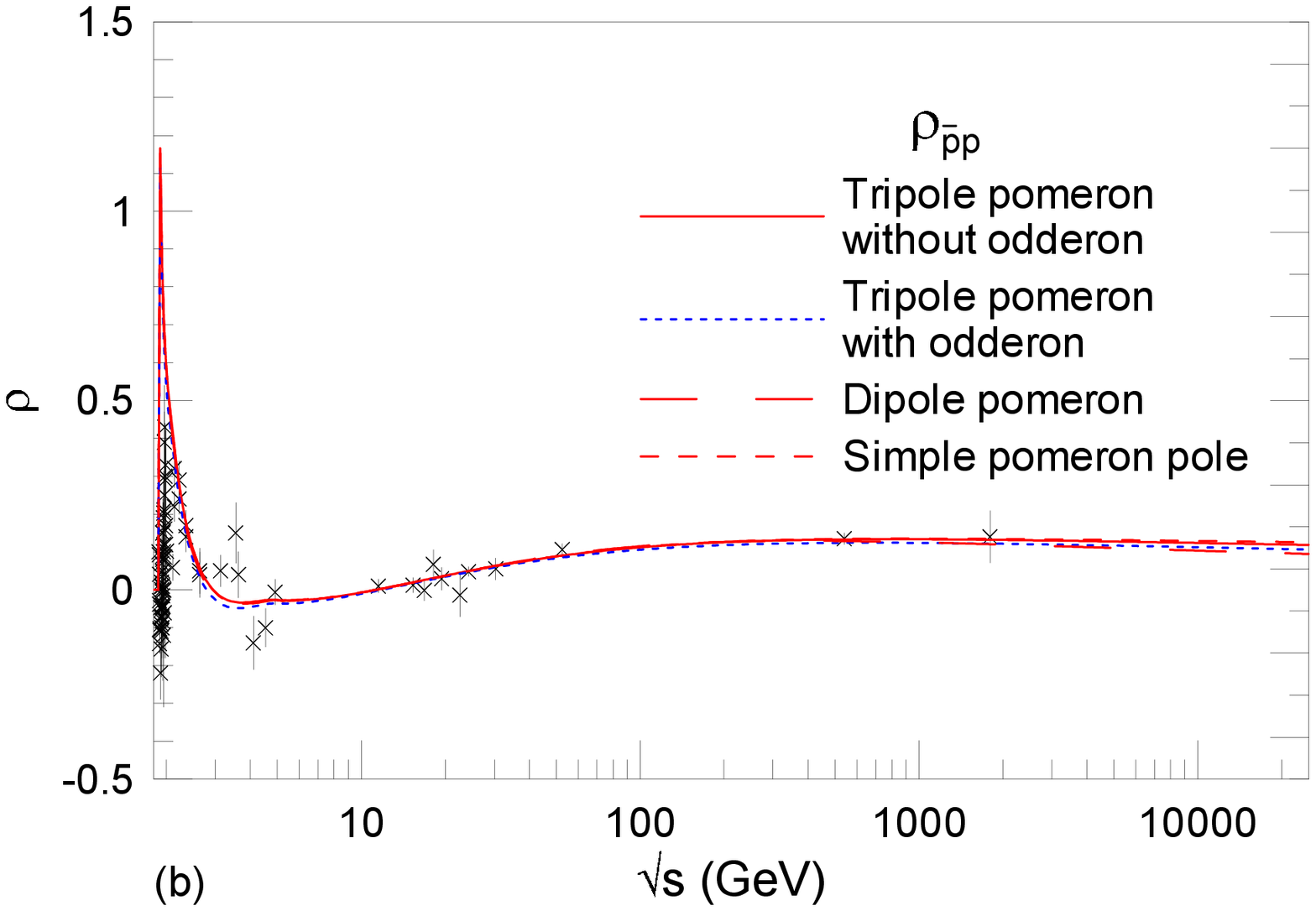}
 \end{minipage}
\caption{Ratio of real part to imaginary part of $pp$ (a) and $\bar pp$ (b) forward scattering amplitudes}
\label{fig:rho-pp-pap}
\end{figure}

\begin{figure}[ht]
\begin{minipage}{7.6cm}
  \includegraphics[width=1.0\textwidth]{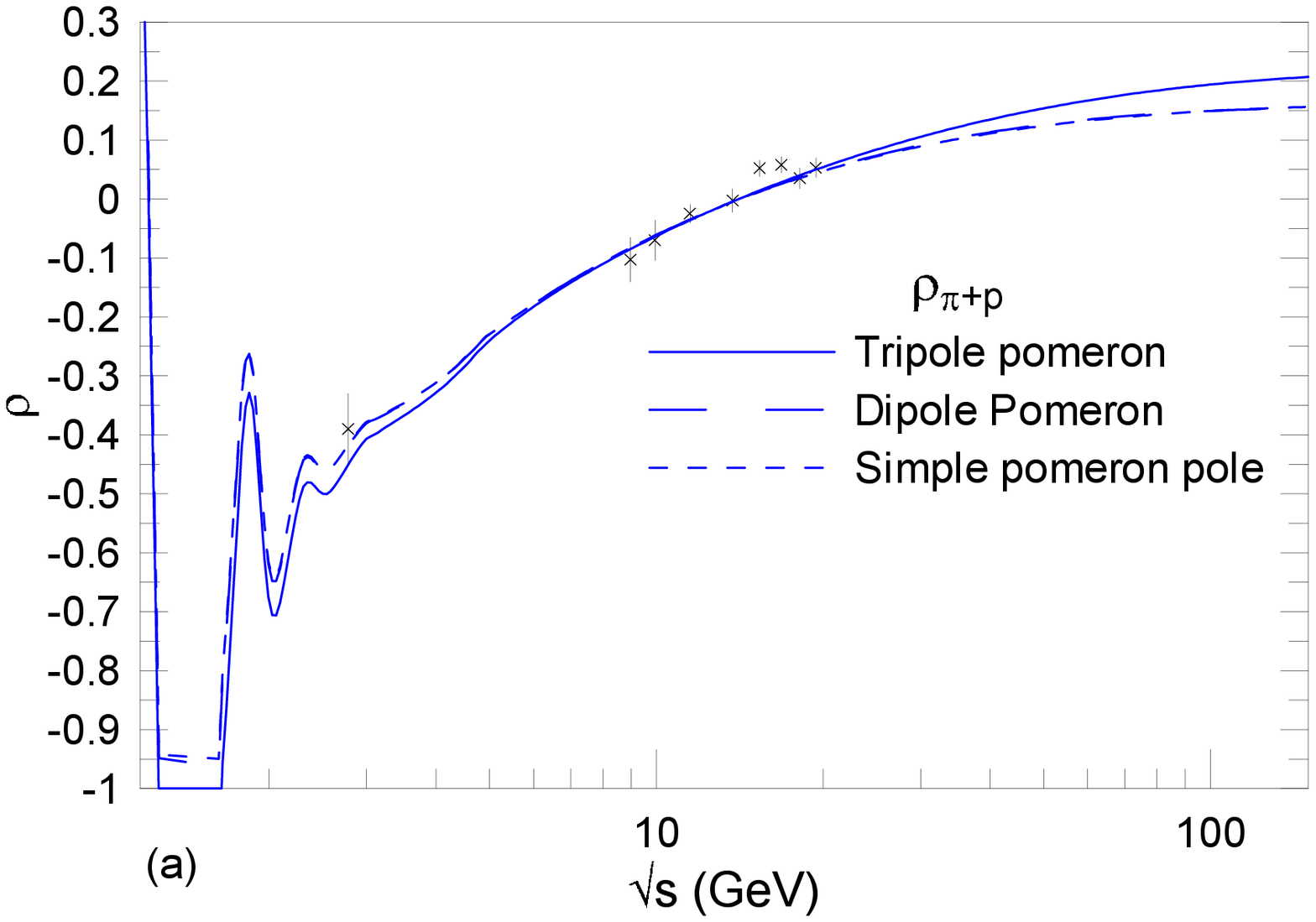}
 \end{minipage}
\begin{minipage}{7.6cm}
  \includegraphics[width=1.0\textwidth]{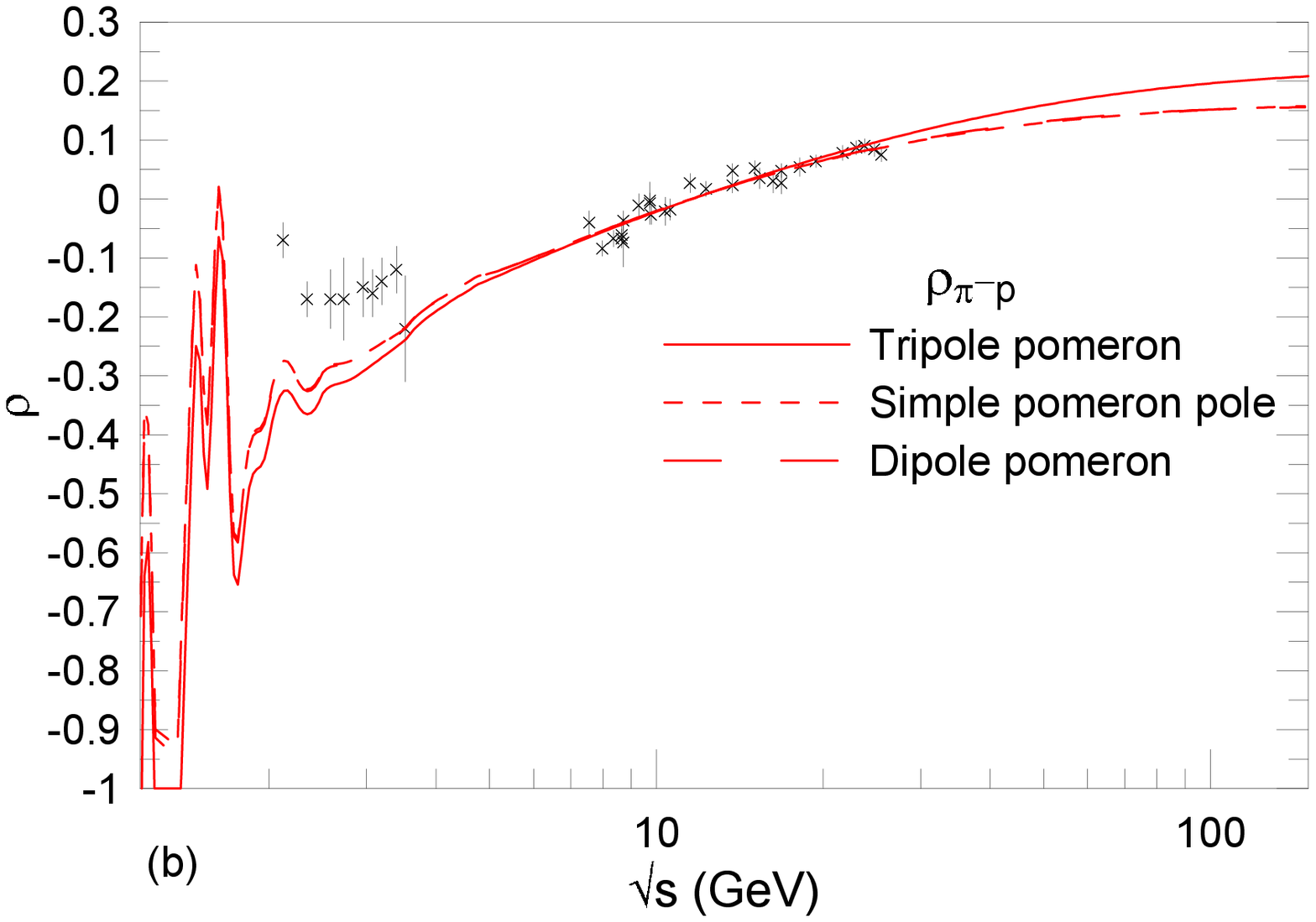}
 \end{minipage}
\caption{Ratio of real part to imaginary part of $\pi^{+}p$ (a) and $\pi^{-}p$ (b) forward scattering amplitudes}
\label{fig:rho-pip-pim}
\end{figure}

\begin{figure}[ht]
\begin{minipage}{7.6cm}
  \includegraphics[width=1.0\textwidth]{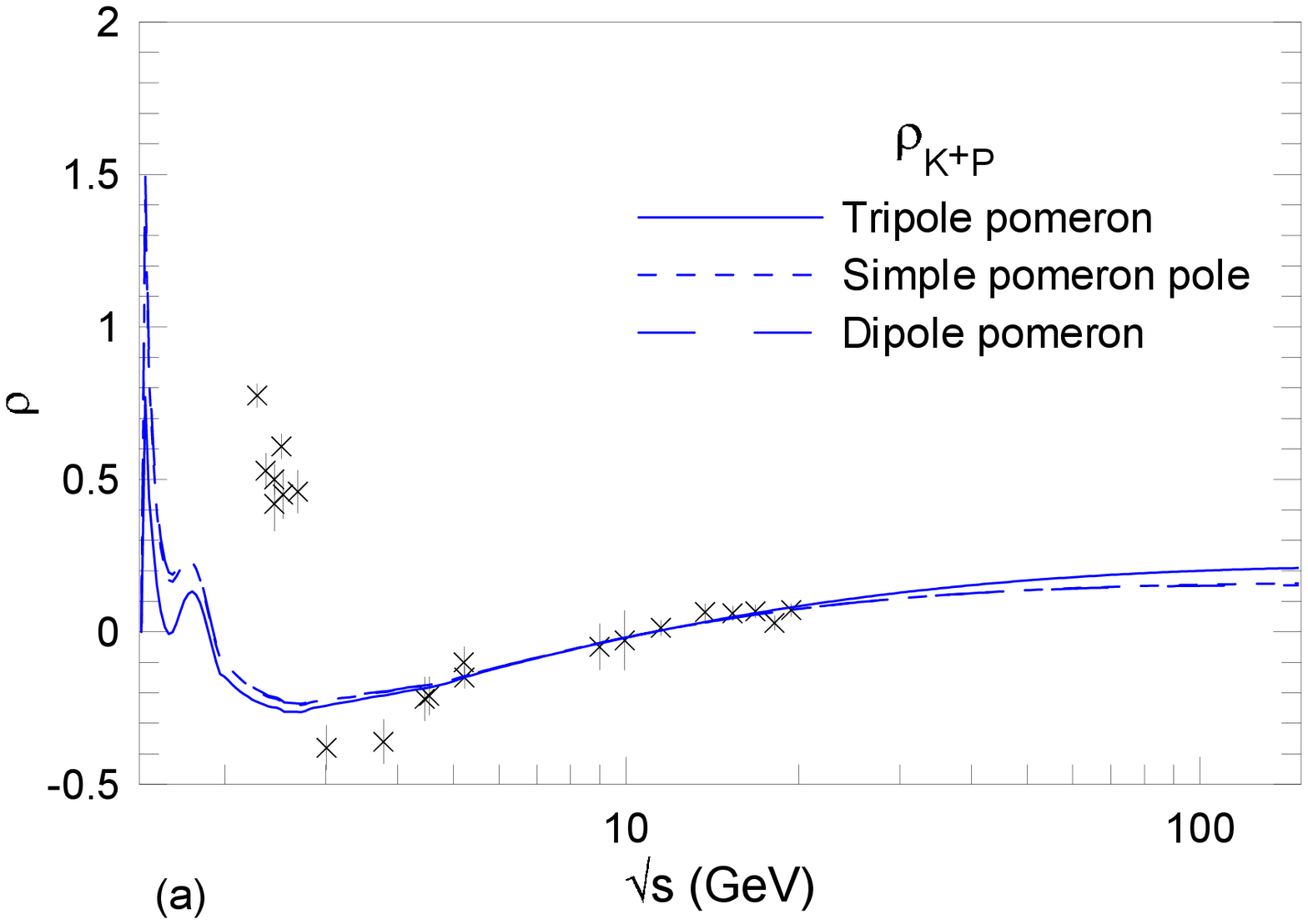}
 \end{minipage}
\begin{minipage}{7.6cm}
  \includegraphics[width=1.0\textwidth]{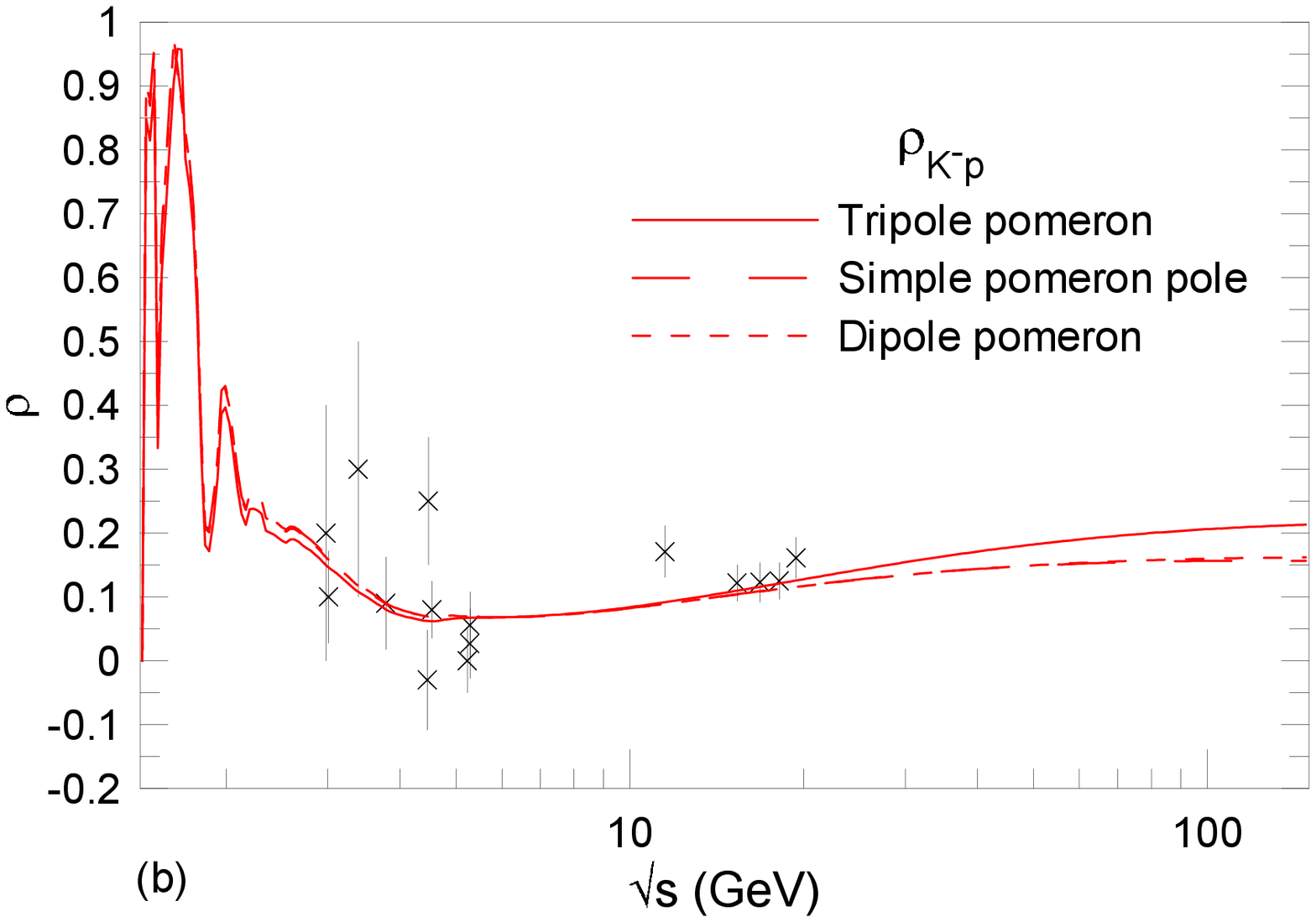}
 \end{minipage}
\caption{Ratio of real part to imaginary part of  $K^{+}p$ (a) and $K^{-}p$ (b) forward scattering amplitudes}
\label{fig:rho-kp-km}
\end{figure}

The quality of the data description in terms of $\chi^{2}/dof$ are again given in the Table
\ref{tab:chi2}, the parameters of the models are presented at the Table
\ref{tab:param}, comparison of the theoretical curves and the data are in the
Figs. \ref{fig:sig-pp-pap}, \ref{fig:sig-pip-pim}, \ref{fig:sig-kp-km} for total cross sections and
in the Figs. \ref{fig:rho-pp-pap}, \ref{fig:rho-pip-pim}, \ref{fig:rho-kp-km} for the ratios of real to imaginary part.


\begin{table}
\caption{The values of parameters obtained in the IDR fit for three pomeron models.}
\label{tab:param}       
\begin{tabular}{lcccccc}
\hline\noalign{\smallskip}
&  \multicolumn{2}{c}{Simple pole pomeron}&\multicolumn{2}{c}{Dipole pomeron}&\multicolumn{2}{c}{Triple pomeron}\\
\cline{2-7}  
&value&error&value&error&value&error\\
\hline\noalign{\smallskip}
Pomeron  &&&&&\\
$\alpha_{\cal P}(0)$  &1.06435 & 0.00027    &--     & --&--     &--\\
$g_{0p}$              &-130.8  &1.1         &-136.7 &1.2 &122.00 &0.23\\
$g_{1p}$              &185.25  &0.70        &31.09  &0.12 &1.146 &0.042\\
$g_{2p}$              &-- &--     &-- &-- &0.9619 &0.0060\\
$g_{0\pi}$            &-20.659 &0.095    &-24.98 &0.10 & 21.554 &0.018\\
$g_{1\pi}$            &20.184 &0.060    &3.607 &0.010 &-2.6941 &0.0029\\
$g_{2\pi}$            & -- &--     & -- & -- &0.25396 &0.00041\\
$g_{0K}$              &-59.91 &0.40    &-49.67 &0.43 &60.51 &0.30\\
$g_{1K}$              &66.32 &0.25    &10.231 &0.049 &-7.178 &0.030\\
$g_{2K}$              &-- &--     &-- &-- &0.8618 &0.0028\\
$R^{+}$-Reggeon       &&&&&\\
$\alpha_{+}(0)$       &0.7012 &0.0018    &0.80483 &0.00069 &0.6244 &0.0011\\
$g_{+p}$              &254.2 &1.7    &432.6 &1.7 &196.66 &0.87\\
$g_{+\pi}$            &41.22 &0.31    &62.05 &0.19 &13.11 &0.12\\
$g_{+K}$              &65.29 &0.68    &118.72 &0.64 &0. &490. \\
$R^{-}$-Reggeon       &&&&&\\
 $\alpha_{-}(0)$       &0.4726 &0.0081    &0.4734 &0.0081 &0.4725 &0.0028\\
$g_{-p}$              &103.3 &3.3    &102.9 &3.3 &103.4 &1.3 \\
$g_{-\pi}$            &7.65 &0.36    &7.58 &0.36 &7.51 &0.17\\
$g_{-K}$              &30.6 &1.1    &30.5 &1.1 &30.50 &0.52\\
Subtraction constants &&&&&\\
$B_{p}$               &-167. &33.     &-172. &33. &-164. &30. \\
$B_{\pi}$             &-78. &21.    &-79.& 21.  &-89.  &19.\\
$B_{K}$               &13.  &26.    &14. & 26.  &6.  &24.\\
\noalign{\smallskip}\hline
\end{tabular}
\end{table}

One can see from the Figures that all the data at $\sqrt{s}>$ 5GeV are reproduced very well in all considered pomeron models. Nevertheless, they show (as should be because of the different asymptotic behavior) the different cross sections and ratios for the energies where there are no data yet.  In Table \ref{tab:predict} we compare predictions for the LHC energies obtained in three pomeron models, using three different methods for data analysis.
The first one implements  the IDR for all amplitudes, the second one is the standard ``-is'' fit with the asymptotic value of the flux factor in the optical theorem, and the third considers \cite{AM} only data on $pp$ and $\bar pp$ were analyzed. The values of $\chi^{2}/dof$ are shown as well for all cases.

\begin{table}
\footnotesize{
\caption{Predictions for the LHC energies of the considered pomeron models obtained within three methods of the fit.  SP is the simple pole pomeron model, DP is  the dipole pomeron model, TP is the tripole pomeron model.}
\label{tab:predict}
\begin{tabular}{lccccccccc}
\hline\noalign{\smallskip}
&  \multicolumn{3}{c}{IDR, $a^{\pm} p, a=p,\pi,K $}&  \multicolumn{3}{c}{``-is'', $a^{\pm} p, a=p,\pi,K $}&  \multicolumn{3}{c}{IDR, $pp, \bar pp$}\\
\cline{2-10}\noalign{\smallskip}  
 & SP & DP & TP  & SP & DP & TP & SP & DP & TP \\
\noalign{\smallskip}\hline\noalign{\smallskip}
$\sqrt{s}=7$TeV&&&&&&&&&\\
$\sigma_{t}$ (mb) & 95.04 & 91.10 & 94.14& 94.92 & 90.79 & 94.20&96.36 & 90.40 & 95.07\\
$\rho$ & 0.138 & 0.112 & 0.138  & 0.186& 0.107 & 0.142 & 0.141 & 0.106 & 0.130 \\
\noalign{\smallskip}\hline
$\sqrt{s}=14$TeV&&&&&&&&&\\
$\sigma_{t}$ (mb) & 105.90 & 99.65 & 104.60 &106.20 & 99.68 & 105.10&108.99 & 98.96 & 106.43\\
$\rho$ & 0.135 & 0.105 & 0.135  & 0.192 & 0.100 & 0.137 & 0.140 & 0.099 & 0.126 \\
\noalign{\smallskip}\hline
$\chi^{2}/dof$&0.976&0.974&0.963&1.14&0.998&0.984&1.096&1.103&1.096\\
\end{tabular}
}
\end{table}

Comparing the results obtained for $pp, \bar pp$ only with those for all processes we first note that the values of $\chi^{2}$ in the later fit are lower. Secondly, the predictions of three pomeron models in the later fit are closer  to each other than those obtained from fitting  $pp, \bar pp$ only. It seems, that the addition of the $\pi p$ and $Kp$ data restricts the freedom of the adjustable parameters of the models.

At the same time the curves $\rho$ for $\pi^{\pm}p$ and $K^{\pm}p$ at $\sqrt{s}<5$ GeV deviate significantly  from the data.
We would like to note that at low energies there is a well pronounced resonance structure in the $\pi^{\pm}p$ and $K^{\pm}p$ cross sections. It is well described by the low-energy parametrization (Section \ref{sec:low-energy}), but if the resonances contribute to the amplitudes, then they should be associated with poles of the amplitudes shifted from the real axis in the complex s-plane. Effectively they are taken into
account in the imaginary part of amplitude, i.e. in the cross section. However they should
contribute as well to the real part of amplitude. If we treat these resonances as stable strong
interacting hadrons or as asymptotic states (in terms of S-matrix theory) we should add
the residues of these poles to the IDR. This would lead to additional constants in the expressions
for $ReA(s, 0)$ in Eqs.(\ref{eq:1s-idr}),  (\ref{eq:2s-idr}). Because of the different sets of the resonances for different processes, these constants must be not the same for $A_{a^{+}p}$ and $A_{a^{-}p}$ amplitudes. We tried to add such constants to the IDR but the decrease of $\chi^{2}$  (about 2-3\%) and they do not improve an agreement with low-energy data. To explain these situation one should presumably argue that new branch points cuts, related to the production of the resonances and their subsequent decay into asymptotic particles,
must be exist and must be taken into account in the IDR. These cuts are not important at high energy but they significantly contribute to the real part of amplitude at very low energy. This problem requires further investigation.

\section{Conclusion}
The method of  integral dispersion relations leads to a better description of the high energy data, giving a $\chi^{2}/dof$  lower by a few percents than other methods,  and  to predictions similar to those obtained through other methods.
LHC predictions of the each model from the three considered methods of analysis, we show
the overall intervals for the predicted  $pp$ cross section and $\rho$ ratios.
\begin{itemize}
\item
Simple-pole pomeron model
\footnotesize{
$$
\sigma_{t}=\left \{
\begin{array}{lll}
\,\,\,\, 94.9 - \,\,\,\, 96.4  &\mbox{mb},  &\sqrt{s}=\,\,\,\,{\rm 7\,\,TeV,}\\
105.9 - 109.0  &\mbox{mb},   &\sqrt{s}={\rm14\,\,TeV.}\\
\end{array}
\right .
\!\! \quad \rho=\left \{
\begin{array}{ll}
0.138 - 0.186,   \quad &\sqrt{s}=\,\,\,\,{\rm 7\,\,TeV,}\\
0.135 - 0.192,  &\sqrt{s}={\rm 14\,\,TeV.}\\
\end{array}
\right .
$$
}
\item
Double-pole pomeron model
\footnotesize{
$$
\sigma_{t}=\left \{
\begin{array}{lll}
90.4 - 91.1  &\mbox{mb}, \,\,\,\,\quad &\sqrt{s}=\,\,\,\,{\rm 7\,\,TeV,}\\
99.0 - 99.7  &\mbox{mb}, \,\,\,\,\quad  &\sqrt{s}={\rm 14\,\,TeV.}\\
\end{array}
\right .
\quad \rho=\left \{
\begin{array}{ll}
0.106 - .112,  \quad  &\sqrt{s}=\,\,\,\,{\rm 7\,\,TeV,}\\
0.10 - 0.11,  \quad &\sqrt{s}={\rm 14\,\,TeV.}\\
\end{array}
\right .
$$
}
\item
Triple-pole pomeron model
\footnotesize{
$$
\sigma_{t}=\left \{
\begin{array}{lll}
\,\,\,\, 94.1 - \,\,\,\,95.1  &\mbox{mb},  &\sqrt{s}=\,\,\,\,{\rm 7\,\,TeV,}\\
104.6 - 106.4  &\mbox{mb},   &\sqrt{s}={\rm 14\,\,TeV,}\\
\end{array}
\right .
\quad \rho=\left \{
\begin{array}{ll}
0.130 - 0 .142,  \quad  &\sqrt{s}=\,\,\,\,{\rm 7\,\,TeV,}\\
0.126 - 0.137,   &\sqrt{s}={\rm 14\,\,TeV.}\\
\end{array}
\right .
$$
}
\end{itemize}
If the precision of the TOTEM measurement of the total $pp$ cross section turns out to be better than 1\% we will
have a chance to select either DP model or SP and TP models (it seems from the total cross
section only it would difficult to distinguish SP and TP models).

\section*{\label{sec:ack}Acknowledgements}
E.M. would like to thank AGO department of the Liege University for the invitation to the
Spa Conference as well as BELSPO for support for his visit to the Liege University where this work has been
completed.

\end{document}